\documentclass[aps,prl,twocolumn,superscriptaddress,showpacs,amsmath,amssymb]{revtex4-1}
\usepackage{graphicx,bbm}
\usepackage{color}

\newcommand{\be}{\begin{equation}}
\newcommand{\ee}{\end{equation}}
\newcommand{\bea}{\begin{eqnarray}}
\newcommand{\eea}{\end{eqnarray}}

\newcommand{\un}[1]{{\underline{#1}}}
\newcommand{\bra}[1]{{\langle #1 \vert}}
\newcommand{\K}{\kappa}

\newcommand{\ket}[1]{{\vert #1 \rangle}}

\newcommand{\braket}[2]{\langle #1 \vert #2 \rangle}

\newcommand{\ave}[1]{{\langle #1\rangle}}
\newcommand{\ii}{ {\rm i} }
\newcommand{\dd}{ {\rm d} }
\newcommand{\NN}{\mathbb{N}}
\newcommand{\ZZ}{\mathbb{Z}}
\newcommand{\RR}{\mathbb{R}}
\newcommand{\CC}{\mathbb{C}}

\newcommand{\y}{{\rm y}}
\newcommand{\x}{{\rm x}}
\newcommand{\z}{{\rm z}}

\newcommand{\mm}[1]{{\mathbf{#1}}}
\def\vmbb#1{\mathbb{#1}}

\def\tr{{\,{\rm tr}}}

\def\one{\mathbbm{1}}

\def\be{\begin{equation}}
\def\ee{\end{equation}}

\def\sp#1#2{\vec{\mm{s}}_{{\rm a}_#1}\cdot \vec{\mm{s}}_{{\rm a}_#2}}

\def\tr{\,{\rm tr}\,}
\def\ket#1{|#1\rangle}
\def\bra#1{\langle#1|}
\def\braket#1#2{\langle #1 | #2 \rangle}

\def\ave#1{\langle #1 \rangle}
\def\ii{{\rm i}}
\def\z{{\rm z}}

\def\tit#1{}
\newcommand{\half}{{\textstyle\frac{1}{2}}}
\newcommand{\nhalf}{{\frac{1}{2}}}
\newcommand{\nthalf}{{\frac{3}{2}}}
\newcommand{\nfhalf}{{\frac{5}{2}}}
\newcommand{\thalf}{{\textstyle\frac{3}{2}}}
\newcommand{\quart}{{\textstyle\frac{1}{4}}}
\newcommand{\ihalf}{{\textstyle\frac{\ii}{2}}}

\begin{document}
\title{Quasilocal conserved operators in isotropic Heisenberg spin 1/2 chain}
\author {Enej Ilievski}
\affiliation{Institute for Theoretical Physics, University of Amsterdam, Science Park 904,
Postbus 94485, 1090 GL Amsterdam, The Netherlands}
\author{Marko Medenjak}
\affiliation{Faculty of Mathematics and Physics, University of Ljubljana, Jadranska 19, SI-1000 Ljubljana, Slovenia }
\author{Toma\v z Prosen}
\affiliation{Faculty of Mathematics and Physics, University of Ljubljana, Jadranska 19, SI-1000 Ljubljana, Slovenia }
\begin{abstract}
Composing higher auxiliary--spin transfer matrices and their derivatives, we construct a family of quasilocal conserved 
operators of isotropic Heisenberg spin 1/2 chain and rigorously establish their linear independence from the well-known set of 
local conserved charges.
\end{abstract}

\maketitle

{\em Introduction.--} The Heisenberg chain of $n$ spins $\half$ with the Hamiltonian (known as the $XXX$ model)
\begin{equation}
H = \sum_{x=0}^{n-1} (\vec{\sigma}_x \cdot \vec{\sigma}_{x+1}+\one),
\label{H}
\end{equation}
where $\vec{\sigma}_x = (\sigma^{\x}_x,\sigma^\y_x,\sigma^\z_x)$ are Pauli operators and periodic boundaries are assumed $\vec{\sigma}_n \equiv \vec{\sigma}_0$,
is arguably the simplest nontrivial interacting quantum many-body model. The spectrum and eigenstates of $H$ can be formulated in terms of the famous Bethe ansatz \cite{bethe}, which gave birth to the theory of quantum integrable systems \cite{faddeev,korepin}.
Eq. (\ref{H}) has been originally proposed as the model of (anti)ferromagnetism in solids \cite{heisenberg} and is indeed a very good description of the modern spin-chain materials \cite{experiments}. It may also be considered as a fundamental paradigm of quantum statistical mechanics which is being used for developing theoretical mechanisms of 
non-equilibrium dynamics and thermalization or relaxation to the generalized Gibbs ensemble (GGE) \cite{silva,GGE_old,GGE}. 

The relaxation dynamics based on quantum quenches \cite{GGEcaux,GGEmarcin,GGEhungary,GGEandrei} gave firm evidence that the full set of ($\sim n$) local conserved operators, the existence of which is 
granted for a quantum integrable system, is {\em incomplete}, in the sense that it cannot describe the steady state completely through a GGE. Similarly, a numerical experiment counting the number of linearly independent time-averaged local operators \cite{MPP15} indicated that the set of local conserved charges should be incomplete and numerical approximations of new quasilocal operators have been put forward. 

In this Letter we explicitly construct new families of non-local but quasilocal operators by composition of a transfer matrix (TM) -- in the sense of algebraic Bethe ansatz, but for higher integer or half-integer auxiliary spins $s > \half$ -- and its derivative, at a special combination of spectral parameters, which in the thermodynamic limit (TL) becomes equivalent to a logarithmic derivative of TM.
Furthermore, we prove quasilocality (in full rigour for a finite set of auxiliary spins $s$) as well as linear independence of these new operator families w.r.t. local conserved charges. Generally, we identify quasilocality with the condition of {\em factorizability} of the largest eigenvalue of an auxiliary TM that enters in the computation of the norm of the conserved operator, i.e. a product of higher-spin TM and its derivative.
As we facilitate finite-dimensional {\em unitary} representations of quantum or Lie symmetries, the new quasilocal operators are always spin-reversal symmetric unlike in alternative recent constructions in the $XXZ$ chain \cite{P11,PI13,P14,affleck} which only work at special, commensurate values of anisotropy. This features promise that our technique shall be applicable for generating new quasilocal charges in other generic integrable models with Lie or quantum group symmetries. Being able to construct a complete, or as large as possible set of independent quasilocal conserved charges is crucial for any application in quantum statistical mechanics, besides constructing GGE, e.g. in linear response theory at finite temperatures, studies of quantum ergodicity and many-body localization, etc. Quasilocal conservation laws are also closely related to boundary driven/dissipative quantum chains \cite{P11,P15}.

{\em Transfer matrices and conserved operators.--} 
Let ${\cal V}_s$, $s\in \half\ZZ^+$, denote a $2s+1$ dimensional spin-$s$ module, ${\cal V}_{s} \equiv \CC^{2s+1} = {\rm lsp}\{ \ket{m},m=-s,-s+1,\ldots,s\}$, ${\rm lsp}$ denoting a linear span of a set of vectors, carrying the {\em unitary} irreducible representation of $SU(2)$ with  generators
\be
\mm{s}^\z\ket{m} = m\ket{m},\; \mm{s}^\pm \ket{m} = \sqrt{(s+1\pm m)(s\mp m)}\ket{m\pm 1}.
\ee
The physical Hilbert space is an $n-$fold tensor product of fundamental representations ${\cal H}_{\rm p} = {\cal V}_{1/2}^{\otimes n}$, 
with $\sigma^\z \equiv 2\mm{s}^\z$, 
$\sigma^\pm = \half(\sigma^\x \pm\ii\sigma^\y) \equiv \mm{s}^\pm$. Fixing arbitrary $s\in \half\ZZ^+$ and 
considering another, {\em auxiliary} Hilbert space ${\cal H}_{\rm a} = {\cal V}_s$, we define Lax matrices as operators over
${\cal H}_{\rm p}\otimes{\cal H}_{\rm a}$
\be
\mm{L}_{x,{\rm a}}(\lambda) =  \lambda\one + \sigma^\z_x \mm{s}^\z_{\rm a} +  \sigma^+_x \mm{s}^-_{\rm a}
+  \sigma^-_x \mm{s}^+_{\rm a} =  \lambda\one  + \vec{\sigma}_x \cdot \vec{\mm{s}}_{\rm a},
\ee
where $\lambda\in\CC$ is the spectral parameter. Throughout the Letter, operators acting nontrivially over the auxiliary Hilbert space are written in bold, or double strike font if acting over multiple (tensor product of) auxiliary spaces.
As a simple consequence of Yang-Baxter equation, the (physical) TMs $T_s(\lambda)\in {\rm End}({\cal H}_{\rm p})$
\be
T_s(\lambda) = \tr_{\!\rm a} \mm{L}_{0,{\rm a}}(\lambda) \mm{L}_{1,{\rm a}}(\lambda) \cdots \mm{L}_{n-1,{\rm a}}(\lambda),
\ee
where $s$ is the auxiliary spin, 
form a commuting family 
\be[T_s(\lambda),T_{s'}(\lambda')] = 0,\quad\forall s,s',\lambda,\lambda'.
\label{comm}
\ee
The fundamenal TM $T_{\nhalf}(\lambda)$ is generating the complete set of local conserved Hermitian operators
\be
Q_k = -\ii\partial_t^{k-1} \log T_{\nhalf}(\half+\ii t)|_{t=0} = \sum_{x=0}^{n-1} \hat{\cal S}^x \left( \one_{2^{n-k}} \otimes q_k \right),
\label{Qk}
\ee
$k\ge 2$, with $Q_2=H$, where $q_k \in {\rm End}({\cal V}^{\otimes k}_{1/2})$ is a $k-$point operator density, and $\hat{\cal S}$ is a cyclic lattice shift map over
${\rm End}({\cal H}_{\rm p})$ defined by $\hat{\cal S}(\sigma^\alpha_{x}) = \sigma^\alpha_{{\rm mod}(x+1,n)}$.

{\em Locality and quasilocality.--}
The $4^n$-dimensional space of physical operators  ${\rm End}({\cal H}_{\rm p})$ is turned into a Hilbert space by defining a Hilbert-Schmidt (HS) inner product
$(A,B) := \ave{A^\dagger B}$ w.r.t. the infinite-temperature state $\ave{A}:=2^{-n}\tr A$.
Let $\{A\} := A - \ave{A}\one$ denote the traceless part of an operator. One of physically most important features of the local conservation laws $Q_k$ is the extensivity of the HS norm
$\| \{Q_k\} \|^2_{\rm HS}:=(\{Q_k\},\{Q_k\}) = \left(2^{-k}\tr(q^\dagger_k q_k) - |2^{-k}\tr q_k|^2\right)n \propto n$. We define (equivalently to \cite{P14}) a general traceless, translationally invariant operator $A = \hat{\cal S}(A) \in {\rm End}({\cal H}_{\rm p})$ as {\em quasilocal} if two conditions are met: (i) $\| A\|^2_{\rm HS} \propto n, $ and (ii) for any locally supported $k$-site operator $b = b_k \otimes \one_{2^{n-k}}$ the overlap $(b,A)$ is asymptotically, as $n\to\infty$, {\em independent} of $n$. 

One should stress that quasilocality only makes sense in TL $n\to\infty$ as it is the property of an infinite sequence of operators labelled by $n$, rather than operators for any fixed size $n$.
More intuitively, a quasilocal operator $Q$ can be thought of as a convergent sum of local operators $Q = \sum_{r=1}^\infty Q^{(r)}$, where $Q^{(r)}$ includes only terms supported on $r$ contiguous sites and the sum $\|Q\|_{\rm HS}^2 = \sum_{r=1}^\infty \| Q^{(r)}\|_{\rm HS}^2$ is rapidly (typically exponentially) converging. Usually \cite{P11,PI13,P14}, quasilocality can be detected by inspecting the leading eigenvalue of a certain
auxiliary transfer matrices, whose $r$-th power yields the partial norm of the $r-$site terms $\| Q^{(r)}\|_{\rm HS}^2$. The effect of quasilocal conserved operators to statistical mechanics is arguably \textit{as important} as that of local operators. In particular, quasilocal charges can be understood as those conserved operators of one-dimensional systems which can influence equilibrated (steady-state) values, say after a quantum quench, of strictly local observables. Our central result is the following 

\smallskip
\noindent
{\bf Theorem:} {\em Traceless operators $X_s(t)$, $s \in\half\ZZ^+,t\in\RR$, defined over the physical Hilbert space ${\cal H}_{\rm p}$ as
\bea
X_s(t) &=& \left[\tau_s(t)\right]^{-n} \left\{ T_s(-\half+\ii t) T'_s(\half + \ii t)\right\}, \label{X} \\
\tau_s(t) &=& -t^2 - \left( s + \half \right)^2, \label{tau}
\eea
where $T'_s(\lambda) \equiv \partial_\lambda T_s(\lambda)$,  are quasilocal for all $s,t$ and linearly independent from $\{Q_k;k\ge 2\}$ for $s > \half$.}

\smallskip
\noindent

The fact that $X_s(t)$ are exactly conserved and $[X_s(t),X_{s'}(t')] = [X_s(t),Q_k]= 0$ follows directly from (\ref{comm}).
The form of our ansatz (\ref{X}) is inspired from an observation (see Eq.~(\ref{Qk}) or, e.g., Ref.~\cite{fagotti}) that at $s=\half$, TM becomes in TL $n\to\infty$ a unitary operator
\be
T_{\nhalf}(\half + \ii t) \simeq \exp\left(\ii \sum_{k=1}^\infty \frac{t^k}{k!}Q_{k+1}\right),
\label{fagotti}
\ee and hence (\ref{X}) can be associated with a logarithmic derivative via 
$T^{\dagger}_{s}(\lambda) \equiv T^T_s(\bar{\lambda}) = (-1)^n T_s(-\bar{\lambda})$ where the last equality is due to spin reversal symmetry $\mm{s}^\z \to -\mm{s}^\z$, $\mm{s}^\pm \to -\mm{s}^\mp$. 

{\em Proof of quasilocality.--} First, we write a matrix product form of a general product of a pair of TMs \cite{katsura}
\be
T_s(\mu)T_s(\lambda) = \tr_{\!\rm a_1,a_2} \prod_{x=0}^{n-1}\left(\sum_{\alpha\in{\cal J}} \vmbb{L}_s^\alpha(\mu,\lambda) \sigma^\alpha_x\right) \\
\ee
where the operators $\vmbb{L}_s^\alpha(\mu,\lambda)$, $\alpha\in{\cal J}:=\{0,\x,\y,\z\}$ act over a pair of auxiliary spaces ${\cal H}_{{\rm a}_1}\otimes {\cal H}_{{\rm a}_2} \equiv {\cal V}_s\otimes {\cal V}_s$
\bea
\vmbb{L}_s^0(\mu,\lambda) &=& \lambda\mu \one + \vec{\mm{s}}_{\rm a_1}\cdot \vec{\mm{s}}_{\rm a_2},\\
\vec{\vmbb{L}}_s(\mu,\lambda) &=& \ii\,\vec{\mm{s}}_{\rm a_1}\!\times\vec{\mm{s}}_{\rm a_2} + \lambda \vec{\mm{s}}_{\rm a_1} + \mu \vec{\mm{s}}_{\rm a_2}.
\eea
Identity component can be written with the Casimir operator $\mm{C}=(\vec{\mm{s}}_{\rm a_1}+\vec{\mm{s}}_{\rm a_2})^2$ as
$\vmbb{L}_s^0 = \mu\lambda \one + \half (\mm{C} - \vec{\mm{s}}_{\rm a_1}^{\, 2} - \vec{\mm{s}}_{\rm a_2}^{\, 2})$, 
hence its spectrum reads $\tau^j_s(\mu,\lambda) = \frac{j(j+1)}{2} - s(s+1) + \mu\lambda$, $j=0,1,\ldots,2s$. Placing the spectral parameters along one of the two lines
\bea
&& {\cal D}^\pm =\{(\mu^\pm_t,\lambda^\pm_t);t\in\RR\} \subset \CC^2, \nonumber\\ 
&& \mu^\pm_t := \mp\half+\ii t,\;\; \lambda^\pm_t :=\pm\half+\ii t,
\eea
we define the restricted auxiliary operators as $\vmbb{L}_s^{\pm\alpha}(t):=\vmbb{L}_s^\alpha(\mu^\pm_t,\lambda^\pm_t)$.
The {\em dominating} eigenvalue of Hermitian operator $\vmbb{L}_s^{+0}(t)\equiv \vmbb{L}_s^{-0}(t)$ is
$\tau_s(t)=\tau^0_s(\mu^\pm_t,\lambda^\pm_t)$, Eq.~(\ref{tau}), corresponding to the {\em singlet} eigenstate
\be
\ket{\psi_0} = (2s+1)^{-1/2}\sum_{m=-s}^s (-1)^{s-m} \ket{m}\otimes\ket{-m},
\ee
with a finite gap to the subleading eigenvalue $\tau'_s(t)$,  $\delta = \log|\tau_s(t)/\tau'_s(t)| > 0$, for any $t$.
The condition $(\vec{\mm{s}}_{\rm a_1}+\vec{\mm{s}}_{\rm a_2})\ket{\psi_0}=0$ and the $SU(2)$ algebra $\vec{\mm{s}}_{{\rm a}_k}\!\times \vec{\mm{s}}_{{\rm a}_k}=\ii \vec{\mm{s}}_{{\rm a}_k}$ imply
the following useful identities
\bea
&& \vec{\vmbb{L}}^+_s(t) \ket{\psi_0} = 0,\quad \bra{\psi_0}\vec{\vmbb{L}}^+_s(t) = -2\bra{\psi_0}\vec{\mm{s}}_{\rm a_1}, \nonumber  \\
&& \bra{\psi_0}\vec{\vmbb{L}}^-_s(t) = 0, \quad \vec{\vmbb{L}}^-_s(t)\ket{\psi_0} = -2\vec{\mm{s}}_{\rm a_1}\ket{\psi_0}.
\label{hit}
\eea
We proceed by constructing a TM over a 4-spin auxiliary space ${\cal H}_{\rm a}=\bigotimes_{k=1}^4 {\cal H}_{\rm a_k}$, ${\cal H}_{\rm a_{1,2}}\equiv {\cal V}_s$, ${\cal H}_{\rm a_{3,4}}\equiv {\cal V}_{s'}$
\be
\vmbb{T}_{s,s'}(\mu,\lambda,\mu',\lambda') =  \sum_{\alpha \in {\cal J}} \vmbb{L}^{\alpha}_s(\mu,\lambda) \otimes \vmbb{L}^\alpha_{s'}(\mu',\lambda'), 
\label{ATM}
\ee
which helps us to compute a general inner product of the form $K_{s,s'}(t,t'):=(X_s(t),X_{s'}(t'))$.
The Hilbert--Schmidt kernel (HSK) then immediately follows after differentiating traces of powers of suitable TMs
\begin{widetext}
\be
K_{s,s'}(t,t') =
[\tau_s(t)\tau_{s'}(t')]^{-n}
\partial_{\lambda^-_t}\partial_{\lambda^+_{t'}}\left(
 \tr \left[\vmbb{T}_{s,s'}(\mu^-_t,\lambda^-_t,\mu^+_{t'},\lambda^+_{t'})\right]^n -
 \tr \left[\vmbb{L}^0_{s}(\mu^-_t,\lambda^-_t)\right]^n \tr \left[\vmbb{L}^0_{s'}(\mu^+_{t'},\lambda^+_{t'})\right]^n
\right).
\label{HSK}
\ee
As a consequence of boundary condition given by Eq.~(\ref{hit}) we obtain that
$\tau_{s,s'}(t,t'):=\tau_s(t)\tau_{s'}(t)$ is always an eigenvalue of $\vmbb{T}_{s,s'}(t,t'):=\vmbb{T}_{s,s'}(\mu^-_t,\lambda^-_t,\mu^+_{t'},\lambda^+_{t'})$ with a product-singlet eigenvector $\ket{\Psi_0} = \ket{\psi_0}\otimes\ket{\psi_0}$. 
One can further show that it is always a dominating and non-degenerate eigenvalue by demonstrating that $\vmbb{T}_{s,s'}(t,t') - \tau_s(t)\tau_{s'}(t)\one$ is a negative definite operator on ${\cal H}_{\rm a} \setminus \CC\ket{\Psi_0}$ 
(see Sects.~A,B of \cite{sup} for details. We note though that we managed to rigorously prove negativity only for $s \le s_0=3/2$ and further succeeded to confirm it with exact numerical computations up to much larger maximal auxiliary spin $s_0$, while for {\em any} $s$ it formally remains a conjecture).
Denoting by $\tau_{s,s'}(\mu,\lambda,\mu',\lambda')$ the continuation of the dominating eigenvalue in the proximity of the domain ${\cal D}^-\!\times {\cal D}^+$, and using Hellmann-Feynman theorem to evaluate its first derivatives
$
\partial_{\lambda^-_t} \tau_{s,s'}(\mu^-_t,\lambda^-_t,\mu^+_{t'},\lambda^+_{t'}) = \partial_{\lambda^-_t} \tau^0_s(\mu^-_t,\lambda^-_t) \tau^0_{s'}(\mu^+_{t'},\lambda^+_{t'})
$,
$
\partial_{\lambda^+_{t'}} \tau_{s,s'}(\mu^-_t,\lambda^-_t,\mu^+_{t'},\lambda^+_{t'}) = \tau^0_s(\mu^-_t,\lambda^-_t)\partial_{\lambda^+_{t'}}\tau^0_{s'}(\mu^+_{t'},\lambda^+_{t'})
$, the HSK can be computed as 
\be
K_{s,s'}(t,t') = n [\tau_{s}(t)\tau_{s'}(t')]^{-1}
\partial_{\lambda^-_t}\partial_{\lambda^+_{t'}}\left(
\tau_{s,s'}(\mu^-_t,\lambda^-_t,\mu^+_{t'},\lambda^+_{t'}) - \tau^0_{s}(\mu^-_t,\lambda^-_t)\tau^0_{s'}(\mu^+_{t'},\lambda^+_{t'})
\right) + {\cal O}(e^{-\gamma n}).
\label{HSKresult}
\ee
\end{widetext}
Remarkably, $n^2$ term exactly cancels, while the finite-size corrections are exponentially small in the gap $\gamma = \log|\tau_{s,s'}(t,t')/\tau'| > 0$ to subleading eigenvalue of $\tau'$ of $\vmbb{T}_{s,s'}(t,t')$.
We shall later derive an explicit expression for HSK. 

What remains to be shown is that $X_s(t)$ have well defined expansions in terms of local operators in TL $n\to\infty$.
For any $k$-local basis operator $\sigma^{\un{\alpha}}_{1:k}:=\sigma^{\alpha_1}_1\sigma^{\alpha_2}_2\cdots \sigma^{\alpha_k}_k$, $\alpha_{1,k}\neq 0$,
we write the component of (\ref{X}) as 
$  [\tau_s(t)]^{-n}\partial_{\lambda^+_t} (\sigma^{\un{\alpha}}_{1:k},T_s(\mu^+_t)T_s(\lambda^+_t))$. 
For treating $n\to\infty$ asymptotics we substitute $[\vmbb{L}_s^{+0}(t)/\tau_s(t)]^{n-k} = \ket{\psi_0}\!\bra{\psi_0} + {\cal O}(e^{-\delta n})$ and take into account 
the fact that the $\lambda-$derivative should always hit the last, $k-$th factor, producing $\partial_\lambda \vec{\vmbb{L}}_s = \vec{\mm{s}}_{\rm a_1}$,  otherwise the whole term
would vanish due to the Eqs.~(\ref{hit}). Thus we find a compact matrix product formula for the components (with the $k=1$ component vanishing)
\be
(\sigma^{\un{\alpha}}_{1:k},X_s(t)) = \bra{\psi_{\alpha_1}} \vmbb{X}^{\alpha_2} \cdots \vmbb{X}^{\alpha_{k-1}}\ket{\psi_{\alpha_k}}
+ {\cal O}(e^{-\delta n}),
\label{MPA}
\ee
where $\vmbb{X}^\alpha := \vmbb{L}_s^{+\alpha}(t)/\tau_s(t)$, $\ket{\psi_\alpha} := \sqrt{2} \ii \mm{s}^\alpha_{\rm a_1}\ket{\psi_0}/\tau_s(t)$.
The HS norm of $X_s(t)$ projected onto $\ell$ sites, in the limit $n-\ell\to\infty$, can be written analogously to Eq.~(\ref{HSK})
\bea
&&\lim_{n\to\infty}\sum_{k=2}^\ell (\ell -k + 1)\sum_{\un{\alpha}}\left\vert(\sigma^{\un{\alpha}}_{1:k},X_s(t))\right\vert^2= \label{HSKx}\\
&&\frac{1}{[\tau_s(t)]^{2\ell}}
\partial_{\lambda^-_t}\partial_{\lambda^+_{t}}\biggl(
 \bra{\Psi_0}\left[\vmbb{T}_{s,s}(\mu^-_t,\lambda^-_t,\mu^+_{t},\lambda^+_{t})\right]^\ell\ket{\Psi_0}  \nonumber \\
&& \qquad -\bra{\psi_0}\left[\vmbb{L}^0_{s}(\mu^-_t,\lambda^-_t)\right]^\ell\ket{\psi_0}\bra{\psi_0}\left[\vmbb{L}^0_{s}(\mu^+_t,\lambda^+_{t})\right]^\ell\ket{\psi_0}
\biggr). \nonumber
\eea
thus resulting in expression $\propto \ell$, cf. Eq.~(\ref{HSKresult}), without any finite-size ($\ell$-dependent) corrections as $\ket{\Psi_0}$ is an exact eigenstate.
We have thus shown that the expansion
\be
X_s(t) = \lim_{\ell\to\infty}\lim_{n\to\infty}\sum_{k=2}^\ell\sum_{\un{\alpha}} (\sigma^{\un{\alpha}}_{1:k},X_s(t)) \sum_{x=0}^{n-1}  \hat{\cal S}^x(\sigma^{\un{\alpha}}_{1:k}),
\label{complete}
\ee
is complete in the HS norm. Q.E.D.

Eqs.~(\ref{HSKx},\ref{complete}) have two useful implications: (i) As the state $\ket{\Psi_0}$ is a spin singlet (in $4-$spin auxiliary space) the only relevant part of the $SU(2)$ invariant TM 
$\vmbb{T}_{s,s'}(t,t')=\sum_\alpha \vmbb{L}^{-\alpha}_s(t) \otimes \vmbb{L}^{+\alpha}_{s'}(t')$, 
is the $(2J+1)$-dimensional block, $J={\rm min}\{s,s'\}$, 
constituting the spin singlet subspace of ${\cal H}_{\rm a}$, where it can be written explicitly as a tridiagonal matrix (see Sect~A of \cite{sup}). (ii) The HSK can be compactly written in terms of the resolvent of the TM, similarly as in \cite{P14},
namely 
$K_{s,s'}(t,t')= n \sum_{k=0}^\infty \bra{\Psi}\bigl[\widetilde{\vmbb{T}}_{s,s'}(t,t')\bigr]^{k}\ket{\Psi}$, where $\widetilde{\vmbb{T}}_{s,s'}(t,t') = \vmbb{T}_{s,s'}(t,t')/[\tau_{s}(t)\tau_{s'}(t')]$ and $\ket{\Psi}=\sum_{\alpha\in\{\x,\y,\z\}} \ket{\psi_{\alpha}}\otimes\ket{\psi_{\alpha}}$, e.g. via solving a system of $2J$ linear equations
\be
K_{s,s'}(t,t') = n \braket{\Psi}{\Phi},\quad (\one - \widetilde{\vmbb{T}}_{s,s'}(t,t'))\ket{\Phi} = \ket{\Psi}.
\label{HSKsystem}
\ee
By deriving the explicit form of matrix elements of $\vmbb{T}_{s,s'}(t,t')$ and solving Eq.~(\ref{HSKsystem}), we can encode the HSK explicitly in terms of
a superposition of Cauchy-Lorentz distributions (assuming $s \le s'$) [see Sect.~A of \cite{sup}]
\bea
&&K_{s,s'}(t,t') = n\frac{ \K_{s,s'}(t-t')}{\tau_{s}(t)\tau_{s'}(t')}, \label{Kex}\\
&&\kappa_{s,s'} = \sum_{l=1}^{2s} \frac{l (l+2(s'\!-\!s))(2s+1-l)(2s'+1+l)}{(2s+1)(2s'+1)}c_{s'-s+l}, \nonumber \\
&& {\rm where} \quad c_s(\tau) := \frac{s}{s^2+\tau^2}.\nonumber
\eea
Note that HSK is symmetric $K_{s,s'}(t,t') = K_{s',s}(t',t)$ and strictly positive $K_{s,s'}(t,t') > 0$, $\forall s,s',t,t'$.
  
We would like to remind the reader that in TL $n\to\infty$ the $s=\half$ family $X_{\nhalf}(t)$ is equivalent to the family of local charges $Q_k$, as follows from Eqs. (\ref{X},\ref{fagotti})
\be
X_{\nhalf}(t) = \sum_{k=0}^\infty \frac{t^k}{k!} Q_{k+2},\quad
Q_{k+2} = \partial_t^k X_{\nhalf}(t)|_{t=0}.
\ee
Eq.~(\ref{MPA}) thus generates also a handy explicit matrix product representations of the standard local conservation laws $Q_k$ or their densities $q_k$.

{\em Proof of linear independence.--} 
Let us first show that $X_{1}(t)$ are linearly independent from $X_{\nhalf}(t)$, i.e., from $Q_k$. We define an operator
\be
\widetilde{X}_1(t) = X_1(t) - \int_{-\infty}^\infty \dd t' f_t(t') X_{\nhalf}(t'),
\label{tX}
\ee
where the function $f_{t}(t')$ is determined by minimizing the HS norm
$\| \widetilde{X}_1(t)\|_{\rm HS}^2$, i.e. by the variation
\be
\frac{\delta}{\delta f_{t}(t')} (\widetilde{X}_1(t),\widetilde{X}_1(t)) = 0,
\ee
resulting in the Fredholm equation of the first kind
\be
\int_{-\infty}^\infty \dd t'' K_{\nhalf,\nhalf}(t',t'') f_{t}(t'') = K_{\nhalf,1}(t',t).
\label{fredholm}
\ee 
Using the fact that the kernels (\ref{Kex}) are related to Cauchy-Lorentz distributions $c_s(t)$ up to trivial rescalings, we make an ansatz $f_t(t') = (\tau_{\nhalf}(t')/\tau_{1}(t)) \varphi(t-t')$ which maps (\ref{fredholm}) to a linear convolution equation 
$\frac{3}{4}c_1 \ast \varphi = \frac{4}{3} c_{\nthalf}$, which, using the well-known convolution identity $c_s \ast c_{s'} = \pi c_{s+s'}$, results in $\varphi=\frac{16}{9\pi}c_{\nhalf}$, or
\be
f_{t}(t') = \frac{8}{9\pi} \frac{1+{t'}^2}{((3/2)^2+t^2)((1/2)^2+(t-t')^2)}.
\label{f0}
\ee
The conclusion of this analysis is that a family $\widetilde{X}_1(t)$ is (a) {\em quasilocal} (see Sect. D of \cite{sup} for a numerical example) as its HSK, computed via Eqs.~(\ref{Kex},\ref{tX},\ref{f0}), is extensive 
$(\widetilde{X}_1(t),\widetilde{X}_1(t')) = \frac{n}{\tau_{1}(t)\tau_1(t')} (\frac{8}{9}c_1(t-t')-\frac{4}{27}c_2(t-t'))$ 
and (b) is orthogonal to (and hence linearly independent from) all known local operators contained in the $s=1/2$ family
$X_{1/2}(t)$, namely we have
$(\widetilde{X}_1(t),Q_k) = (\widetilde{X}_1(t),X_{\nhalf}(t'))=0$, for all $t,t',k$. 
More generally, one can orthogonalize $X_s(t)$ for higher $s$ to all previous $X_{s'}(t')$ for $s'<s$,
by making an ansatz $\widetilde{X}_s(t) = X_s(t) - \int_{-\infty}^\infty\dd t'\left(f^t_{s,s-\nhalf}(t')X_{s-\nhalf}(t') + 
f^t_{s,s-1}(t')X_{s-1}(t')\right)$, with explicit expressions for bounded integrable functions 
$f^t_{s,s-1}(t'),f^t_{s,s-\nhalf}(t')$. These families are HS orthogonal for different auxiliary spins, namely $(\widetilde{X}_s(t),\widetilde{X}_{s'}(t')) = 0$ for $s\neq s'$, while $(\widetilde{X}_s(t),\widetilde{X}_{s}(t)) > 0$, i.e. $\widetilde{X}_s(t)\neq 0$, for all $s,t$ (for details see Sect.~C of \cite{sup}). This implies that $X_s(t)$ are linearly independent from all previous $X_{s'}(t')$, for $s'<s$, and in particular from $X_{\nhalf}(t')$ or $Q_k$. Q.E.D.

{\em Discussion.--} We have proposed a direct extension of local conserved operators derived from the logarithm of the fundamental TM \cite{faddeev,korepin,grabowski} to higher spin auxiliary spaces. We have proved that in such a case, the resulting operators are quasilocal. An interesting side-result of our statement is an 
asymptotic (thermodynamic), $n\to\infty$, inversion formula \cite{Maillard} $T^{-1}_s(\half + \ii t) \simeq [\tau_s(t)]^{-1} T_s(-\half + \ii t)$, valid for any $s\in\half\ZZ^+$, which can be proven by implementing our matrix product formula (\ref{MPA}) together with the gap
statements (Sect. B of \cite{sup}) to show that $T_s(\mu^\pm_t)T_s(\lambda^\pm_t) \simeq \tau_s(t) \one$.
Our quasilocal operators $X_s(t)$ (\ref{X}) can thus be understood as {\em logarithmic derivatives} of 
$T_s(\lambda^+_t)$. In TL $n\to\infty$ they become Hermitian operators for any $t\in\RR$. For $s=\half$, the Taylor expansion coefficients in $t$ turn out to be {\em local} operators, while for $s > \half$, they remain non-local but {\em quasilocal}.
One could thus equivalently work with a discrete
series of quasilocal operators $Q_{s,k+2} = (1/k!) \partial_t^k X_s(t)|_{t=0}$, $s\in\half\ZZ^+,k\in\ZZ^+$, rather than with a series of continuous families $X_s(t)$. As a double index suggests, the number of relevant quasi-local charges in a large finite system may grow as $n^2$, rather than $n$ as in the ultralocal case, although this question cannot be made precise with the results at hand.

Our results promise a number of timely applications and generalizations. The new quasilocal families should be included in order to correctly describe $k\to 0,\omega\to 0$ limit of dynamical structure factors and general Drude weights at finite temperatures \cite{zotos,robin,IP13}, or GGE in quantum quench protocols \cite{GGE}. For computing stationary expectations of local observables after a quench from a non-thermal initial state, such as e.g. the N\'eel state $\ket{{\rm N}}$, one can readily demonstrate extensivity $\bra{{\rm N}}X_s(t)\ket{{\rm N}} \propto n$ by extracting the leading eigenvalue of an associated transfer matrix, essentially proceeding along the lines of calculation done in Ref.~\cite{fagotti} for the fundamental ($s=1/2$) TM.
Appropriate $q-$deformations of the concepts developed in this Letter should provide additional quasilocal operator families for the anisotropic Heisenberg model ($XXZ$ chain).
Extensions to $SU(N)$ symmetric integrable spin chains seem straightforward, whereas a generalization to continuous quantum integrable systems and field theories (such as Lieb-Liniger or sine-Gordon models) should be a challenge for the future.  We close by stressing an important point of distinction with respect to spin-reversal symmetry breaking quasilocal
conserved operators in $XXZ$ model \cite{P11,PI13,P14,affleck}. Quasilocality, as abstractly formulated here, requires a finite-dimensional (but non-fundamental) representation of a quantum TM, and a factorizability condition for the leading eigenvalue of the associated auxiliary TM.
This can happen, either for irreducible unitary representations of the symmetry group, but will result in operators which are always even under spin reversal, as is the case here; or due to the root-of-unity (commensurability) condition for the anisotropy, where highest-weight type non-unitary representations become reducible to finite dimensional ones, such as in the $XXZ$ model.

E.~I. thanks J. de Nardis and J.-S. Caux while T.~P. thanks M.~Mierzejewski and P.~Prelov\v sek for valuable discussions on related topics.
M.~M. and T.~P. acknowledge support by Slovenian Research Agency (grants P1-0044, J1-5439 and N1-0025).

\setcounter{equation}{0}
\makeatletter
\renewcommand{\theequation}{S\arabic{equation}}
\renewcommand{\thefigure}{S\arabic{figure}}
\renewcommand{\bibnumfmt}[1]{[S#1]}
\renewcommand{\citenumfont}[1]{S#1}

\newpage
\begin{widetext}

\pagebreak

\begin{center}
\textbf{\large Supplemental material:\\ Quasilocal conserved operators in isotropic Heisenberg spin 1/2 chain}
\end{center}

\end{widetext}

\section{A: Explicit computation of the Hilbert-Schmidt kernel}

Let us define the following operator over the $4-$spin auxiliary space ${\cal H}_{\rm a}=\bigotimes_{k=1}^4 {\cal H}_{{\rm a}_k}$:
\be
\vmbb{F}_{s,s'}(t,t'):=\vmbb{T}_{s,s'}(t,t')-\tau_{s}(t)\tau_{s'}(t')\one
\label{defF}
\ee 
where $\vmbb{T}_{s,s'}(t,t')$ has been defined in the main text. For $s=s'$ and $t=t'$, the operator $\vmbb{T}_{s,s}(t,t)$ is the TM for computation of the HS norm, so its maximal eigenvalue has to be positive as the HS norm is always positive. 
To complete the proof of quasilocality, it amounts to show that $\vmbb{F}_{s,s}(t,t)$ is negative definite on ${\cal H}_{\rm a} \setminus \CC\ket{\Psi_0}$ where $\ket{\Psi_0}=\ket{\psi_0}\otimes\ket{\psi_0}$ is the
leading eigenvector, which implies that $[\tau_{s}(t)]^2$ is the maximal and non-degenerate eigenvalue with a {\em finite} gap to the sub-leading eigenvalue.
This will be shown by explicitly constructing the operator (\ref{defF}) in a convenient basis (see section B for completing the proof of the gap), 
which in turn is needed for explicitly computing the HSK
$K_{s,s'}(t,t') = (X_s(t),X_{s'}(t'))$ via Eq.~(\ref{HSKsystem}) of the main text.
 
Using the expansion
\be
\vmbb{F}_{s,s'}(t,t')=\sum_{a,b\in\{0,1,2\}}^{a+b\le 2}t^{a}t'^{b}\;\vmbb{F}_{a,b},
\ee
and suppressing dependence  $s,s'$ in $\vmbb{F}_{a,b}$ for compactness of notation, the non-vanishing matrix-components can be readily represented in terms of $SU(2)$ invariant tensors $\vmbb{F}_{a,b}$ by employing a shorthand notation for elementary $SU(2)$ symmetric operators over ${\cal H}_{\rm a}$: $[\![i,j]\!]:=\sp{i}{j}$ and
$[\![i,j,k]\!]:=\ii(\vec{\mm{s}}_{{\rm a}_i}\times\vec{\mm{s}}_{{\rm a}_j})\cdot \vec{\mm{s}}_{{\rm a}_k}$, namely:
\begin{widetext}
\bea
&&
\vmbb{F}_{0,0}=-\quart s(s+1)-\quart s'(s'+1)-s(s+1)s'(s'+1) 
+ [\![1,2]\!][\![3,4]\!] - [\![1,3]\!][\![2,4]\!] + [\![1,4]\!][\![2,3]\!] \nonumber \\
&&\qquad + \half \left( [\![1,2,3]\!] - [\![1,2,4]\!] - [\![1,3,4]\!] + [\![2,3,4]\!] \right) 
+ \quart\left( [\![1,4]\!] + [\![2,3]\!] - [\![1,2]\!] - [\![1,3]\!] - [\![2,4]\!] - [\![3,4]\!]\right),
\\
&&
\vmbb{F}_{1,0}=\ihalf\left( [\![1,3]\!] + [\![2,3]\!] - [\![1,4]\!] - [\![2,4]\!] \right) + \ii\left([\![1,3,4]\!] + [\![2,3,4]\!]\right),
\\
&&
\vmbb{F}_{0,1}=\ihalf\left( [\![2,3]\!] + [\![2,4]\!] - [\![1,3]\!] - [\![1,4]\!] \right) + \ii\left([\![1,2,3]\!] + [\![1,2,4]\!]\right),
\\
&&
\vmbb{F}_{1,1} = -[\![1,3]\!] - [\![1,4]\!] - [\![2,3]\!] - [\![2,4]\!],
\\
&&
\vmbb{F}_{0,2} = -s(s+1) - [\![1,2]\!],
\\
&&
\vmbb{F}_{2,0} = -s'(s'+1) - [\![3,4]\!]. \label{defFend}
\eea

\subsection{Reduction to the singlet subspace}
By virtue of $SU(2)$ invariance of $\vmbb{T}_{s,s'}(t,t')$ the computation of HSK can be facilitated (assuming $s\leq s'$ throughout this section without
loss of generality) in the invariant $2s+1$ dimensional subspace formed by singlet eigenstates,
${\cal H}_{\rm a}\supset {\cal V}_0={\rm lsp}\{\ket{j};j=0,1,\ldots,2s\}$, which can be expanded explicitly in terms of computational 
basis $\ket{m_{1}m_{2}m'_{1}m'_{2}} = \ket{m_1}_{{\rm a}_1}\ket{m_2}_{{\rm a}_2} \ket{m_1'}_{{\rm a}_3}\ket{m_2'}_{{\rm a}_4}$ with help of Wigner $3j$-symbols
\be
\ket{j}=\sum_{M=-j}^{j}(-1)^{j-M}\sqrt{2j+1}\sum_{m_{1},m_{2}=-s}^{s}
\begin{pmatrix}
s & s & j \cr
m_{1} & m_{2} & -M
\end{pmatrix}
\sum_{m'_{1},m'_{2}=-s'}^{s'}
\begin{pmatrix}
s' & s' & j \cr
m'_{1} & m'_{2} & M
\end{pmatrix}\ket{m_{1}m_{2}m'_{1}m'_{2}}, \label{j}
\ee
\end{widetext}
where only the extremal singlet state factorizes $\ket{0}\equiv \ket{\Psi_0} = \ket{\psi_0}_{{\rm a}_1{\rm a}_2}\ket{\psi_0}_{{\rm a}_3{\rm a}_4}$.
Let us denote the restriction to the singlet subspace as
$\vmbb{F}^{(0)}_{s,s'}(t,t'):=\vmbb{F}_{s,s'}(t,t')|_{{\cal V}_{0}}$.

The proof is based on showing the following elementary statements: (i) $\vmbb{F}^{(0)}_{s,s'}(t,t')$ is a quadratic form
in the difference variable $\tau:=t-t^{'}$, with identically vanishing linear terms
$\vmbb{F}^{(0)}_{1,0}=\vmbb{F}^{(0)}_{0,1}\equiv 0$, i.e.
\be
\vmbb{F}^{(0)}_{s,s'}(t,t')\equiv \vmbb{F}^{(0)}_{s,s'}(\tau)=\vmbb{D}\;\tau^{2}+\vmbb{F}^{(0)}_{0,0},
\ee
where $\vmbb{D}=\vmbb{F}^{(0)}_{0,2}=\vmbb{F}^{(0)}_{2,0}=-\half \vmbb{F}^{(0)}_{1,1}$,\\ 
(ii) $\vmbb{F}^{(0)}_{s,s}(t,t')$ is
a real symmetric tridiagonal matrix in orthonormal singlet basis $\{\ket{j}\}$.

Both statements follow from demonstrating, by employing certain elementary symmetry-based reductions, that all matrices
$\vmbb{F}^{(0)}_{a,b}$ are expressible solely in terms of knowing only five types of matrix elements, namely
\bea
&&\bra{j}[\![1,3]\!]\ket{j}, \\ 
&&\bra{j}[\![1,3]\!]\ket{j+1},\\
&&\bra{j}[\![1,3]\!][\![2,4]\!]\ket{j+1},\\
&&\bra{j}[\![1,2,3]\!]\ket{j+1},\quad
\bra{j}[\![2,3,4]\!]\ket{j+1},
\label{cubic_identity}
\eea
for all $\ket{j}\in {\cal 
V}_0$.
The reductions of above expressions can be carried out after noticing simple transformation properties of $\ket{j}$ 
with respect to permutation operators $P_{12}$ and $P_{34}$ ($P_{ij} \in {\rm End}({\cal H}_{\rm a})$ swaps ${\cal H}_{{\rm a}_i}$ and ${\cal H}_{{\rm a}_j}$)
\be
P_{12}\ket{j}=(-1)^{j+2s}\ket{j},\quad P_{34}\ket{j}=(-1)^{j+2s'}\ket{j}.
\label{permutations}
\ee
We find straightforward implications
\bea
&&
\bra{j}[\![1,4]\!]\ket{j'} = \bra{j}[\![2,3]\!]\ket{j'},
\label{ident1}
\\
&&
\bra{j}[\![1,3]\!]\ket{j'}=\bra{j}[\![2,4]\!]\ket{j'}
\label{ident2}
\\
&&
\bra{j}[\![1,2,3]\!]\ket{j'} = -\bra{j}[\![1,2,4]\!]\ket{j'},
\label{ident3}
\\
&&
\bra{j}[\![1,3,4]\!]\ket{j'} = -\bra{j}[\![2,3,4]\!]\ket{j'}.
\label{ident4}
\eea
for any $j,j'$, where, in addition, a sign reversal under odd permutation of factors has been used for the triple-product terms.
Notably, these identification are enough to establish vanishing of linear terms, $\vmbb{F}^{(0)}_{1,0}=\vmbb{F}^{(0)}_{0,1}\equiv 0$.
Furthermore, with an aid of Casimir invariants
\bea
&&
[\![1,2]\!]\ket{j}=\left(-s(s+1)+\half j(j+1)\right)\ket{j}, \\
&&
[\![3,4]\!]\ket{j}=\left(-s'(s'+1)+\half j(j+1)\right)\ket{j},
\label{Casimirs}
\eea
we have $\bra{j}\vmbb{F}^{(0)}_{0,2}\ket{j}=\bra{j}\vmbb{F}^{(0)}_{2,0}\ket{j}=-\half j(j+1)$.
With assistance of symbolic algebra in {\em Mathematica}, using explicit form of singlet eigenstates (\ref{j}) for general $s,s'$ and $j$,
we have been able to obtain $\bra{j}\vmbb{F}^{(0)}_{1,1}\ket{j}=j(j+1)$, whereas vanishing of the upper-diagonal
follows from Eqs.~(\ref{ident1})-(\ref{ident4}). The $\tau^{2}$-dependence (i) is at the end given by
elements $\bra{j}\vmbb{F}^{(0)}_{0,2}\ket{j}=2\bra{j}[\![1,3]\!]\ket{j}=-\half j(j+1)$.

Turning attention to the remaining (constant) term $\vmbb{F}^{(0)}_{0,0}$ we first note that all triple-product terms vanish on
${\cal V}_0$ after resorting to explicit evaluation of matrix elements from Eq.~(\ref{cubic_identity})
\begin{widetext}
\be
\bra{j}[\![1,2,3]\!]\ket{j+1}=-\bra{j}[\![2,3,4]\!]\ket{j+1}=
\frac{(j+1)^{2}}{4}\sqrt{\frac{(2s-j)(2s'-j)(2(s+1)+j)(2(s'+1)+j)}{(2j+1)(2j+3)}}.
\ee
in conjunction with Eqs.~(\ref{ident3},\ref{ident4}). Notably, diagonal matrix elements of all triple-product terms vanish in 
accordance with their skew-symmetric nature.
Furthermore, after repeating symmetry arguments based on Eq.~(\ref{permutations}) we 
arrive at the following explicit expressions: (a) for the diagonal matrix elements $a_j =\bra{j}\vmbb{F}^{(0)}_{0,0}\ket{j}$
\be
a_j=-\quart j(j+1)-s(s+1)s'(s'+1)+
\left(\half j(j+1)-s(s+1)\right)\left(\half j(j+1)-s'(s'+1)\right),
\label{aj}
\ee
and (b) for the first off-diagonal elements $b_j = \bra{j}\vmbb{F}^{(0)}_{0,0}\ket{j+1} = \bra{j+1}\vmbb{F}^{(0)}_{0,0}\ket{j}$
\be
b_j=-\bra{j}[\![1,3]\!]+2[\![1,3]\!][\![2,4]\!]\ket{j+1}=
-\frac{j(j+1)(j+2)}{4}\sqrt{\frac{(2s-j)(2s'-j)(2(s+1)+j)(2(s'+1)+j)}{(2j+1)(2j+3)}}
\label{bj}
\ee

We still owe the reader a brief remark in order to close (ii): by inspecting the structure of singlet eigenstates we can easily see
that $SU(2)$ invariant appearing in the expansion of $\vmbb{T}_{s,s'}(t,t')$ can raise/lower any magnetic quantum numbers
at most by two, thus $\vmbb{F}^{(0)}_{s,s'}(t,t')$ is a banded matrix which cannot have non-vanishing matrix elements beyond the 
second upper/lower diagonals.
At last, due to symmetry cancellations of the second upper diagonal between $[\![1,3]\!][\![2,4]\!]$ and
$[\![1,4]\!][\![2,3]\!]$ projected onto singlets $\ket{j}$, we finally remain with a strictly tridiagonal form.
Finally, $\vmbb{F}^{0}_{s,s'}(t,t')$ is a symmetric matrix on ${\cal V}_0$, i.e. $\bra{j}\vmbb{F}^{0}_{s,s'}
(t,t')\ket{j'}=\bra{j'}\vmbb{F}^{0}_{s,s'}(t,t')\ket{j}$, as consequence of mutual cancellation of skew-symmetric triple-product terms $[\!
[i,j,k]\!]$.

It is useful to note that the matrix of the linear system given by Eq.~(\ref{HSKsystem}) in fact coincides with 
$\vmbb{F}^{(0)}_{s,s'}(\tau)$,
\be
(\one-\widetilde{\vmbb{T}}_{s,s'}(t,t'))=-[\tau_{s}(t)\tau_{s'}(t')]^{-1}\vmbb{F}^{(0)}_{s,s'}(\tau).
\ee
Furthermore, note that the state $\ket{0}$ does not couple to the rest of $2s$-dimensional singlet space ${\cal V}_0'={\rm lsp}\{\ket{j},j=1,\ldots,2s\}$, since $a_0=b_0=0$.
Thus, by solving the $2s$-dimensional tridiagonal linear system
\be
\vmbb{F}^{(0)}_{s,s'}(\tau){\ket{\Xi}}=\ket{1},
\label{trisys}
\ee
we can readily obtain few explicit results for HSKs,
\be
K_{s,s'}(t,t')= n[\tau_{s}(t)\tau_{s'}(t')]^{-1}\kappa_{s,s'}(t-t'),
\quad
{\rm with}\quad
\kappa_{s,s'}(\tau)=-\frac{4}{3}s(s+1)s'(s'+1)\braket{1}{\Xi},
\label{HSKS}
\ee
where the prefactor in front of $\kappa_{s,s'}(\tau)$ comes from expressing the state $\ket{\Psi}$ (see main text), expressed as $\ket{\Psi} = (2/\sqrt{3})\sqrt{s(s+1)s'(s'+1)} \ket{1}$. More explicitly, and using a suitable gauge transformation (redefinition of bra-ket basis in ${\cal V}'_0$ in order to remove the square-roots from off-diagonal matrix elements) and homogenizing the system (\ref{trisys}), one can encode the HSK as
\be
\kappa_{s,s'}(\tau) = -8 s(s+1)s'(s'+1) \frac{\chi_1}{\chi_0},
\label{kappa1}
\ee
where $\chi_j$ satisfies a 3-point recurrence relation 
\be
(j+1)(2s+1+j)(2s'+1+j)\chi_{j-1} + j(2s-j)(2s'-j) \chi_{j+1} + (2j+1)(z-j(j+1)) \chi_j = 0, 
\label{kappa2}
\ee
where $z\equiv 2(\tau^2 + (s+\half)^2+(s'+\half)^2)$,
which can be solved with a direct backward iteration by choosing the initial conditions $\chi_{2s+1}=0$, $\chi_{2s}=1$.

The solution, which is easily obtained explicitly for essentially arbitrary large $s,s'$, can be neatly written in terms of a superposition of Cauchy distributions
\be 
c_s(t) = \frac{s}{s^2 + t^2},
\ee
and for few smallest auxiliary spins reads
\bea
&&
\K_{\nhalf,s} = \frac{2s(s+1)}{2s+1} c_{s+\nhalf}, \nonumber\\
&&
\K_{1,1} = \frac{8}{9} c_1 + \frac{20}{9} c_2,\quad
\K_{1,\nthalf} = \frac{5}{3} c_\nthalf + 3 c_{\nfhalf},\quad
\K_{1,2} = \frac{12}{5} c_{2} + \frac{56}{15} c_3,\quad
\K_{1,\nfhalf} = \frac{28}{9} c_{\nfhalf} + \frac{20}{9} c_{\frac{7}{2}},\quad
\K_{1,3} = \frac{80}{21} c_3 + \frac{36}{7} c_4,\nonumber\\
&&
\K_{\nthalf,\nthalf} = \frac{15}{16} c_1 + 3 c_2 + \frac{63}{16} c_3,\quad
\K_{\nthalf,2} = \frac{9}{5} c_{\nthalf} + \frac{21}{5} c_{\nfhalf} + \frac{24}{5} c_{\frac{7}{2}},\quad
\K_{\nthalf,\nfhalf} = \frac{21}{8} c_2 + \frac{16}{3} c_3 + \frac{45}{8} c_4,\nonumber\\
&&
\K_{2,2} = \frac{24}{25} c_1 + \frac{84}{25}c_2 + \frac{144}{25}c_3 +  \frac{144}{25}c_4,\quad
\K_{2,\nfhalf} = \frac{28}{15}c_{\nthalf} + \frac{24}{5} c_{\nfhalf} + \frac{36}{5} c_{\frac{7}{2}} + \frac{20}{3}c_{\frac{9}{2}}.
\eea
Moreover, a simple form for the superposition coefficients for all small $s,s'$ lead us to conjecture that they can be written as low-order rational expressions. Indeed we found a remarkably simple closed form expression
\be
\kappa_{s,s'}(\tau) = \sum_{l=1}^{2s} \frac{l (l+2(s'\!-\!s))(2s+1-l)(2s'+1+l)}{(2s+1)(2s'+1)}c_{s'-s+l}(\tau),
\label{KexA}
\ee
which reproduces the solution of Eqs.~(\ref{kappa1},\ref{kappa2}) for any finite $s,s'$, while we leave its rigorous derivation for the future.

For a curiosity, we may write another closed form expression of HSKs for general $s,s'$ and $\tau$,
written in terms of a complex continuation of harmonic numbers known as the digamma function $\psi(z)$,
\be
\psi(z)=\frac{\dd}{\dd z}\log \Gamma(z)=\sum_{k=1}^{\infty}\left(\frac{1}{k}-\frac{1}{k+z-1}\right)-\gamma_{\rm EM},
\ee
where $\gamma_{\rm EM}$ is the Euler--Mascheroni constant, as
\be
\kappa_{s,s'}(\tau)=
\frac{s(s'(s'+1)+(s+1)^{2}+\tau^{2})}{2s+1}-
\frac{((s'-s)^{2}+\tau^{2})((s+s'+1)^{2}+\tau^{2})}{2(2s+1)(2s'+1)}
\left(\psi^{+}_{s,s'}+\psi^{-}_{s,s'}\right),
\ee
making use of a compact notation $\psi^{\pm}_{s,s'}:=\psi(s+s'+1\pm \ii \tau)-\psi(s'-s+1\pm\ii \tau)$.
Remarkably, the second term gives rise to Cauchy distributions in (\ref{KexA}) via recurrence formula
\be
\psi(z+N)-\psi(z)=\sum_{k=0}^{N-1}\frac{1}{z+k},\quad N\in \NN,
\ee
yielding
\be
\psi^{+}_{s,s'}+\psi^{-}_{s,s'}=\sum_{k=0}^{2s-1}\left(\frac{1}{k+(s'-s+1+\ii \tau)}+\frac{1}{k+(s'-s+1-\ii \tau)}\right)
=2\sum_{k=1}^{2s}c_{s'-s+k}(\tau).
\ee
 
\section{B: Finiteness of the gap for the auxiliary transfer matrix}

Extensive $\sim n$ scaling of a general HSK can be attributed to the finite spectral gap with respect to the leading-modulus 
eigenvalue of $\vmbb{F}_{s,s'}(t,t')$ on the entire ${\cal H}_{\rm a}$. Thanks to Cauchy--Schwartz inequality
\be
K_{s,s'}(t,t') \le \sqrt{ K_{s,s}(t,t) K_{s',s'}(t',t')},
\ee
it is sufficient to
focus on $s'=s$ and $t'=t$ case only (pertaining to HS norm of $X_{s}(t)$), where we expand
\begin{equation}
\vmbb{F}_{s,s}(t,t)=\vmbb{A}t^{2}+\vmbb{B}t+\vmbb{C},
\label{ABC}
\end{equation}
with matrix-valued coefficients reading
\bea
&&
\vmbb{A}=-2s(s+1)
-([\![1,2]\!] + [\![1,3]\!] + [\![1,4]\!] + [\![2,3]\!] + [\![2,3]\!] + [\![3,4]\!]) \equiv -\half \mm{C}^{[4]}_{s}.
\\
&&
\vmbb{B}=-\ii([\![1,4]\!] - [\![2,3]\!] - [\![1,2,3]\!] - [\![1,2,4]\!] - [\![1,3,4]\!] - [\![2,3,4]\!]),
\\
&&
\vmbb{C}=-\half s(s+1)-s^2(s+1)^{2}
+[\![1,2]\!][\![3,4]\!]-[\![1,3]\!][\![2,4]\!]+[\![1,4]\!][\![2,3]\!]\nonumber
\\
&&
+\quart([\![1,4]\!] + [\![2,3]\!] - [\![1,2]\!] - [\![1,3]\!] - [\![2,4]\!] - [\![3,4]\!])+\half([\![1,2,3]\!] - [\![1,2,4]\!] - [\![1,3,4]\!] + [\![2,3,4]\!]).
\label{C1}
\eea
\end{widetext}
These are just specialization of expressions given by Eqs.(\ref{defF}--\ref{defFend}).
The operator $\mm{C}^{[4]}_{s}:=(\vec{\mm{S}}_{a_1} + \vec{\mm{S}}_{a_2} + \vec{\mm{S}}_{a_3} + \vec{\mm{S}}_{a_4})^{2}$ denotes
the four-fold $s$-spin Casimir invariant with eigenvalues $s(s+1)$. Note that the auxiliary operator denoted by $\CC$, such as in Eqs.~(\ref{ABC},\ref{C1}),  should not be confused with a set of complex numbers.

Denoting temporarily $\vmbb{T}_{s,s}(t,t)\to \vmbb{T}(t)$
we note a remarkable commutativity property,
\be
[\vmbb{T}(t),\vmbb{T}(t')]=0,\quad \forall t,t',
\ee
which is a direct consequence of Yang-Baxter equation. Specifically, considering a periodic chain of four spins $s$ the auxiliary TM $\vmbb{T}(t)$ becomes the standard commuting
quantum TM for the physical spin 1/2 now playing the role of auxiliary spin.
This implies commutativity of all operator valued coefficients, 
\be
[\vmbb{A},\vmbb{B}]=[\vmbb{A},\vmbb{C}]=[\vmbb{B},\vmbb{C}]=0.
\label{ABC2}
\ee
In order to prove strict negativity of $\vmbb{F}_{s,s}(t,t)$ on ${\cal H}_{\rm a}\setminus\CC \ket{\Psi_0}$ it is enough to show that a {\em quadratic (in $t$) equation} $\bra{\Phi}\vmbb{F}_{s,s}(t,t)\ket{\Phi}=0$ does not have a solution, for any  
$\ket{\Phi}$ other than $\ket{\Psi_0}$. Due to (\ref{ABC2}) this amounts to demonstrate that a matrix-valued discriminant
\be
\Delta:=\vmbb{B}^2-4\vmbb{A}\vmbb{C},
\ee
has only \textit{non-positive} eigenvalues, while for any eigenvector $\ket{\Phi_0}$ of $\Delta$ corresponding to zero eigenvalue, it must hold that $\bra{\Phi_0}\vmbb{F}_{s,s}(t,t)\ket{\Phi_0} < 0$.

Indeed, the entire singlet subspace ${\cal V}_{0}$ has the latter property, since we have $\Delta|_{{\cal V}_0} \equiv 0$ 
due to $\vmbb{A}|_{{\cal V}_0}=\vmbb{B}|_{{\cal V}_0}\equiv 0$. 
The negativity of $\vmbb{F}^{(0)}_{s,s}(\tau=0) \equiv \vmbb{F}^{(0)}_{0,0}$ on ${\cal V}'_0$ follows from diagonal dominance of the tridiagonal matrix
\be
-a_j >  |b_j| + |b_{j-1}|,\quad 
j \ge 1,
\ee 
based on explicit form of matrix elements (\ref{aj},\ref{bj}).

Clearly, for large enough $t$ (Casimir) coefficient $\vmbb{A}$ starts to dominate and therefore (non-singlet) eigenstates belonging to
any higher spin multiplet necessarily become sub-leading and the spectral gap $\gamma > 0$ is always due to the largest (smallest in modulus) (singlet) eigenvalue of $\vmbb{F}^{(0)}_{0,0}$. 
For a generic $t\in \RR$ on the other hand it might happen
that the gap $\gamma$ is determined by eigenvectors outside
of ${\cal V}_{0}$. 
At the moment we have only been able to rigorously confirm our statement for $s\in\{ \half,1,\thalf\}$ by 
analytically diagonalizing the operator $\Delta$ projected onto highest-weight total spin $S>0$ subspaces of ${\cal H}_{\rm a}$ ($SU(2)$ descendants only contribute to degeneracies), or 
some larger $s$, by extensive numerical checks.

\section{C: Fredholm-Gram-Schmidt orthogonalization for higher auxiliary spins}

Using the appealing explicit form of HSK (\ref{Kex}), derived in Sect.~A, we here outline a general scheme of orthogonalization of $X_s(t)$ to $X_{s'}(t')$  for all $s'<s$, $t'\in\RR$.
We denote such orthogonalized quasilocal conserved operators as $\widetilde{X}_s(t)$. Picking a set of suitable functions $f^{t}_{s,s'}(t')$, for $s' \in\half\ZZ^+ < s$, we seek for an operator \be
\widetilde{X}_s(t) = X_s(t) - \sum_{s'}^{s'<s} \int_{-\infty}^\infty\!\!\dd t' f^{t}_{s,s'}(t') X_{s'}(t'),
\label{Xt}
\ee
which minimizes the HS norm $\| \widetilde{X}_s(t)\|^2_{\rm HS}$, i.e.,
\be
\frac{\delta}{\delta f^{t}_{s,s'}(t')} (\widetilde{X}_s(t),\widetilde{X}_s(t)) = 0,\quad s' < s.
\ee
This yields a coupled linear system of $(2s-1)\times(2s-1)$ Fredholm equations of the first kind
\bea
&&\sum_{s''}^{s'' < s} \int_{-\infty}^\infty\!\!\dd t'' K_{s',s''}(t',t'') f^{t}_{s,s''}(t'') = K_{s',s}(t',t), \nonumber \\ 
&& \forall t',\, s'<s. \noindent 
\label{FS}
\eea
If the unknown functions $f^t_{s,s'}(t')$ are sought for in terms of the following difference ansatz
\be
\varphi_{s'',s}(t''-t) := \frac{\tau_s(t)}{\tau_{s''}(t'')} f^{t}_{s,s''}(t''),
\ee
then, noting that the HSK (23,\ref{HSKS}) also obeys a scaled difference form
\be
\K_{s',s''}(t'-t'') = n^{-1}\tau_{s'}(t') \tau_{s''}(t'') K_{s',s''}(t',t''),
\ee
the Fredholm system (\ref{FS}) becomes equivalent to a linear convolution system
\be
\sum_{s''}^{s''< s} \K_{s',s''} \ast \varphi_{s'',s} = \K_{s',s},\quad s' < s,
\label{FS2}
\ee
where $(\varphi \ast \varphi')(t):=\int_{-\infty}^\infty \dd t' \varphi(t') \varphi'(t-t')$. 

\medskip 
\noindent
For $s=1$, this yields a single condition
\be
\frac{3}{4}c_1 \ast \varphi_{\nhalf,1} = \frac{4}{3} c_{\nthalf},
\ee
with a unique solution, equivalent to  (\ref{f0}),
\be
\varphi_{\nhalf,1} = \frac{16}{9\pi} c_{\nhalf},
\ee
which follows directly from an elementary addition identity for the Cauchy distributions 
\be
c_s \ast c_{s'} = \pi c_{s+s'}.
\ee

\medskip
\noindent
For $s>1$ the system (\ref{FS}) becomes nontrivial. Then, it turns advantageous to construct a linear isomorphism $\Lambda : \varphi \rightarrow g$ between the convolution ring of 
integrable functions (or distributions) $\varphi(t)$ with operations $(+,\ast)$ spanned by $\{ c_s,s\in\half\ZZ^+\}$, and the ring of functions $g(z)$ of a formal variable $z$ with operations $(+,\cdot)$, where $\cdot$ is the usual pointwise multiplicaiton, analytic on the unit disc around the origin $z=0$. 
The map $\Lambda$ and its inverse $\Lambda^{-1}$ are defined uniquely by:
\bea
&&\Lambda(c_s) = \pi z^{2s}, \qquad\Lambda(\varphi \ast \varphi') = \Lambda(\varphi) \Lambda(\varphi'), \label{Lambda}  \\
&&\Lambda^{-1}(z^k) = \frac{1}{\pi} c_{\frac{k}{2}},\;\;\;\Lambda^{-1}( g g') = \Lambda^{-1}(g)\ast \Lambda^{-1}(g'), \nonumber
\eea
and the linearity.
\noindent
Note that a constant function in the image of $\Lambda$ corresponds to a Dirac distribution $c_0(t) = \frac{1}{\pi}\delta(t)$,
which, however, never appears in our calculation.

Using the following notation for the unknown functions $g_{s',s} := \Lambda(\varphi_{s',s})$, the Fredholm system (\ref{FS}) is $\Lambda$-mapped to $(2s-1)\times(2s-1)$
system of linear equations with coefficients that are polynomials in variable $z$
\be
\sum_{s''}^{s''<s} G_{s',s''}(z) g_{s',s}(z) = G_{s',s}(z),
\ee
where (now assuming $s\le s'$ without loss of generality)
\bea
&& G_{s,s'}(z) := \Lambda(\kappa_{s,s'})(z) = G_{s',s}(z) =\\
&& \pi\sum_{l=1}^{2s} \frac{l (l+2(s'\!-\!s))(2s+1-l)(2s'+1+l)}{(2s+1)(2s'+1)} z^{2(s'-s+l)}. \nonumber 
\eea
Elementary algebra yields a solution which is nonvanishing only for the last two components $s'=s-1$ and $s'=s-\half$ (note that here $s>1$):
\bea
g_{s',s}(z) &=& 0, \quad {\rm for}\; \; s' < s-1,\\
g_{s-1,s}(z) &=& \frac{1}{(s\!-\!1)^2}\left[ \left(1 - \frac{z^2}{\zeta_s}\right)^{\!-1}\!\!\!-1\right] - \frac{s(2s\!-\!1)}{s(2s\!-\!1)-1}z^2, \nonumber \\
g_{s-\nhalf,s}(z) &=& \frac{2(2s)^2}{(2s\!-\!1)(2s\!+\!1)} z
\left[1 - \frac{1}{s(2s\!+\!1)}\left(1-\frac{z^2}{\zeta_s}\right)^{\!\!-1}\right]\nonumber \\
{\rm where} &&\quad \zeta_s = \frac{s(2s + 1)}{(s - 1)(2s - 1)}.
\eea
Note that the convergence radius $\sqrt{\zeta_s}$ is always larger than $1$, guaranteeing analyticity inside the unit disc.
Expanding the geometric series and transforming back with $\Lambda^{-1}$ (\ref{Lambda}), we obtain explicit results
for the two nonvanishing functions 
\bea
f^t_{s,s-1}(t') &=& \frac{\tau_{s-1}(t')}{\pi\tau_{s}(t)} \Biggl( -  \frac{s(2s\!-\!1)}{s(2s\!-\!1)-1}c_1(t'-t)\nonumber \\
&& \qquad + \frac{1}{(s-1)^2}\sum_{l=1}^\infty \zeta_{s}^{-l} c_{l}(t'-t)\Biggr), \nonumber \\
f^t_{s,s-\nhalf}(t') &=&  \frac{2(2s)^2 \tau_{s-\nhalf}(t') }{\pi(2s\!-\!1)(2s\!+\!1) \tau_{s}(t)} \Biggl( c_{\nhalf}(t'-t) \nonumber \\
&&  - \frac{1}{s(2s\!+\!1)} \sum_{l=0}^\infty \zeta^{-l}_s c_{l+\nhalf}(t'-t)\Biggr), \label{explicit}
\eea
which complete the explicit construction of $\widetilde{X}_s(t)$ (\ref{Xt}).
We note that the exponentially convergent sums above, Eqs.~(\ref{explicit}), allow closed form expressions in terms of the Hypergeometric function $_2{F}_1$, or the incomplete Beta function, of argument $1/\zeta_s$ and with complex parameters.

It may be of interest also to consider HS-norms and HS-kernels defined with respect to orthogonalized quasilocal operators
\be
\widetilde{K}_{s}(t,t') = (\widetilde{X}_s(t),\widetilde{X}_{s}(t')) = n\frac{\widetilde{\kappa}_{s}(t-t')}{\tau_s(t)\tau_{s'}(t')}.
\ee
For example, showing that $\widetilde{K}_s(t,t) = \| \widetilde{X}_s(t)\|^2_{\rm HS} > 0$ is a final step of the proof that the $X_s(t)$ are linearly independent for different $s$. In the opposite case, specifically if, for some $s$, $X_s(t)$ would be expressible as a linear combination of $X_{s'}(t')$, for $s'<s$, then one would have $\widetilde{X}_s(t) = 0$, and hence $\widetilde{K}_s(t,t)=0$.

Clearly, designating $\widetilde{G}_{s}=\Lambda(\widetilde{\kappa}_s)$, we find
\bea
\widetilde{G}_s(z) &=& G_{s,s}(z) - \sum_{s',s''}^{s',s''<s} g_{s',s}(z)G_{s',s''}(z)g_{s'',s}(z) \nonumber \\
&=&G_{s,s}(z) - \sum_{s'}^{s'<s} g_{s',s}(z)G_{s',s}(z)   \\
&=&\frac{(2s)^2\pi}{(s-1)^2(2s-1)(2s+1)} \nonumber \\
&&\times\left[ \frac{s(s\!-\!1)(2s\!-\!1)}{2s\!+\!1}z^2 + 1 - \left(1-\frac{z^2}{\zeta_s}\right)^{\!-1}\right], \nonumber
\eea
and transforming back
\bea
&&\widetilde{\kappa}_s(\tau) = 
\frac{(2s)^2}{(s-1)^2(2s-1)(2s+1)} \nonumber \\
&&\times\left[ \frac{s(s\!-\!1)(2s\!-\!1)}{2s\!+\!1}c_1(\tau) - \sum_{l=1}^\infty \zeta^{-l}_s c_{l}(\tau)\right].
\eea
Specifically, noting that $c_{s'}(0) = 1/{s'}$:
\bea
\widetilde{\kappa}(0) &=& \frac{(2s)^2}{\pi(s-1)^2(2s-1)(2s+1)}\Bigl[\frac{s(s\!-\!1)(2s\!-\!1)}{2s\!+\!1}\nonumber\\
&& \qquad  + \log\frac{4s^3-2s+1}{s(2s-1)(2s+1)}\Bigr],
\eea
which satisfies $\widetilde{\kappa}(0) > 0$ for any $s > 1$, and hence $\|\widetilde{X}_s(t)\|_{\rm HS}^2=\widetilde{K}_s(t,t) = 
n\,\widetilde{\kappa}(0)/[\tau_s(t)]^2 > 0$. Note that the case $s=1$ has been treated separately before.

\section{D: Numerical example}

Here we write out, explicitly, the leading terms of the simplest new quasilocal conserved operator that is orthogonal to all the local ones, namely $\widetilde{X}_{s=1}(t=0)$.
Matrix product formula  (\ref{MPA}) inserted to 
Eq.~(\ref{tX}) with (\ref{f0}) yields all the local terms in the infinite size limit $n\to\infty$, say up to 
support size $\ell\le 4$ \bea
&&\widetilde{X}_1(0) = -\frac{7\!\cdot\!2^5}{3^7}\sum_{x=0}^{n-1}\Bigl(\vec{\sigma}_x\cdot\vec{\sigma}_{x+2}
+ \frac{155}{252}\vec{\sigma}_x\cdot\vec{\sigma}_{x+3} \nonumber \\
&&+ \frac{16}{63}(\vec{\sigma}_x\!\cdot\!\vec{\sigma}_{x+1})(\vec{\sigma}_{x+2}\!\cdot\!\vec{\sigma}_{x+3}) -\frac{53}{84}(\vec{\sigma}_x\!\cdot\!\vec{\sigma}_{x+2})(\vec{\sigma}_{x+1}\!\cdot\!\vec{\sigma}_{x+3})
\nonumber\\
&&- \frac{11}{84}(\vec{\sigma}_x\!\cdot\!\vec{\sigma}_{x+3})(\vec{\sigma}_{x+1}\!\cdot\!\vec{\sigma}_{x+2})\Bigr) + {\rm h.o.t.} \label{example}
\eea
We note that this qualitatively agrees with the optimal quasilocal conserved operator $Q'$ which has been constructed approximately by a systematic numerical procedure in Ref.~[1]. 
Small quantitative deviations in the coefficients (note that [1] used spin-1/2 operators rather than Pauli matrices which attributes a relative factor of $4$ in quartic/quadratic terms) can be explained by the 
fact that the operator (\ref{example}), being just one member of the $s=1$ family $\widetilde{X}_1(t)$,
is not optimized with respect to a relative weight within a finite support $\ell$,
$\lambda_\ell(X) = \lim_{n\to\infty}\sum_{k=1}^\ell\sum_{\un{\alpha}}|(\sigma^{\un{\alpha}}_{1:k},X)|^2/(X,X)$.
On the other hand, the operator $Q'$ of [1] is determined precisely by maximizing $\lambda_\ell(X)$
within a given set of conserved $X$. For $\widetilde{X}_1(0)$ we obtain $\lambda_\ell=0.508, 0.682, 0.797$, for $\ell=4,5,6$, respectively, while for the optimal numerical $Q'$ one  has [1] $\lambda_\ell = 0.605, 0.759, 0.840$. Moreover, numerical inspection of relative weights $\mu_\ell(X)=\lambda_\ell(X)-\lambda_{\ell-1}(X)$ of a sequence of higher quasilocal operators, $\widetilde{X}_{s}(0)$, $s=1,\thalf,2$, indicates that 
for larger $s$ the relative weights $\mu_\ell(\widetilde{X}_{s}(0))$ have clear maxima at larger supports $\ell\sim \ell^*(s)$, while after that they decay exponentially $\mu_\ell \sim e^{-\gamma\ell}, \ell > \ell^*(s)$.

\medskip\noindent
$[1]$ M.~Mierzejewski, P. Prelov\v sek, and T. Prosen, Phys. Rev. Lett. {\bf 114}, 140601 (2015).

\end{document}